\documentclass[aps,onecolumn,nofootinbib,floatfix,letterpaper,showpacs,notitlepage]{revtex4-1}

\usepackage{natbib}
\usepackage{times}
\usepackage{graphicx}
\usepackage{epsfig}
\usepackage{subfigure}
\usepackage{amssymb,amsmath}
\usepackage{hyperref}
\usepackage{setspace}
\usepackage{url}

\hypersetup{
    colorlinks = true,
    linkcolor = blue,
    anchorcolor = blue,
    citecolor = blue,
    filecolor = blue,
    pagecolor = blue,
    urlcolor = black
}

\begin{document}

\title{    Muons from Neutralino Annihilations in the Sun: Flipped SU(5)}

\author{Muhammad Adeel Ajaib}
\email{adeel@udel.edu}
\affiliation{Bartol Research Institute, Department of Physics and Astronomy\\
University of Delaware, Newark, Delaware 19716}

\author{Ilia Gogoladze\footnote{On leave of absence from: Andronikashvili Institute of Physics, GAS, Tbilisi, Georgia.}}
\email{ilia@bartol.udel.edu}
\affiliation{Bartol Research Institute, Department of Physics and Astronomy\\
University of Delaware, Newark, Delaware 19716}

\author{Qaisar Shafi}
\email{shafi@bartol.udel.edu}
\affiliation{Bartol Research Institute, Department of Physics and Astronomy\\
University of Delaware, Newark, Delaware 19716}



\begin{abstract}

We consider two classes of supersymmetric flipped SU(5) models with gravity mediated supersymmetry breaking such that the thermal neutralino relic abundance provides the observed dark matter density in the universe. We estimate the muon flux induced by neutrinos that arise from neutralino annihilations in the Sun and discuss prospects for detecting this flux in the IceCube/Deep Core experiment. We also provide comparisons with the corresponding fluxes in the constrained minimal supersymmetric standard model and non-universal Higgs models. Regions in the parameter space that can be explored by the IceCube/DeepCore experiment are identified.

\end{abstract}


\maketitle


\section{Introduction}

It is now widely accepted that approximately 23 $\%$  of the Universe's energy density consists of non-baryonic cold dark matter  \cite{Komatsu:2008hk}. A large number of experiments consisting of direct, indirect  and accelerator searches are currently underway all hoping to discover the underlying, presumably massive ($\sim$ GeV -TeV), weakly interacting dark matter particle (WIMP). The lightest neutralino in supersymmetric models with conserved matter parity is a particularly attractive cold dark matter candidate and has attracted a great deal of attention. The direct detection searches have already yielded important constraints on the spin independent neutralino-nucleon cross sections in the constrained minimal supersymmetric standard model (CMSSM) and some related models (see \cite{Feng:2010gw}, and references therein).

Indirect WIMP searches rely on the capture and subsequent annihilation, say in the Sun's center, of relic dark matter particles. The neutralinos, in particular, can annihilate into the known SM particles, for example, $\chi\chi \rightarrow \tau^+\tau^-$. The tau particles in turn produce energetic muon neutrinos which interact with the polar ice to produce muons which can be identified by the $\rm km^3$ IceCube/Deep Core detector \cite{Collaboration:2010hr}.

Neutralinos in the galactic halo passing through a massive body like the Sun can get captured if they scatter off the nuclei with velocities smaller than the escape velocity. In the core of the Sun, where they eventually accumulate, these neutralinos can annihilate into known SM particles, for e.g., $\chi\chi \rightarrow \tau^+\tau^-$ . These particles decay (e.g. $\tau \rightarrow \nu_\tau \bar{\nu}_\mu\mu$) and produce energetic muon neutrinos which can then be detected at IceCube after they interact with the polar ice and produce muons (for e.g. via processes like $\nu_\mu+N \rightarrow \mu^- +X$, N being the nucleon and X some hadronic system). We investigate the possibility of detecting these energetic neutrinos by estimating the flux of muons that they induce. In addition to the well studied  CMSSM, we explore other well motivated models, namely, flipped SU(5), non-universal Higgs models (NUHM2), and flipped SU(5) with universal Higgs masses at $M_{GUT}$. The prospects for detecting this neutrino induced muon flux by the IceCube/DeepCore experiment is discussed.

In this paper we are mainly interested in studying the implications of supersymmetric flipped SU(5) models for indirect dark matter WIMP searches, with
the lightest neutralino being the dark matter candidate. Flipped SU(5) has several distinct features which are not easily replicated in other GUTs
such as SU(5) and SO(10). For instance, the well-known doublet-triplet splitting problem is easily solved in flipped SU(5) \cite{ant}. Primordial inflation with
predictions for the cosmological parameters in good agreement with the 7 year WMAP data are readily obtained , which in turn, lead to testable predictions for
proton decay \cite{Rehman:2009yj}.

Our paper is organized as follows. In section \ref{framework}, following \cite{Gogoladze:2009mc}, we briefly describe the two flipped SU(5) models under discussion. Consistent with the underlying gauge group (SU(5) x U(1)), both classes of models work with
non-universal gaugino masses. Their difference stems from the non-universal soft scalar Higgs masses employed in one of the models. In section \ref{flux-sd} we review the calculations of the conversion factors relating the muon flux and spin dependent (SD) cross section in the IceCube/Deep Core experiment. Section \ref{procedure} contains a description of the experimental constraints and the scanning procedure employed to generate the benchmark points.  Our predictions for the muon flux  and  SD cross section are presented in section \ref{sec:results}, and  the conclusions are summarized in section \ref{conclusions}.

\section{Theoretical Framework of the Models}\label{framework}

We are interested in estimating the neutrino induced muon fluxes from neutralino annihilations in the Sun, with flipped SU(5) (FSU(5)) boundary conditions imposed on the soft supersymmetry breaking (SSB) parameters at $M_{GUT}$. More generally, we compare four distinct models, namely CMSSM, NUHM2, FSU(5) and FSU(5) with Universal SSB Higgs mass  boundary condition  (FSU(5)-UH). We will briefly describe each model below.
 The CMSSM \cite{Feldman:2008hs} has the following parameters at the $M_{GUT}$:
\begin{equation}
m_0,m_{1/2}, A_0, \rm tan\beta, sign({\mu}).
\label{cmssm-params}
\end{equation}
Here $m_0$ is the soft supersymmetry breaking (SSB) scalar masses, $m_{1/2}$ is the SSB gaugino mass,
$A_0$ is the universal SSB trilinear scalar interaction (with the
corresponding Yukawa coupling factored out), ${\rm tan\beta}$ is the
ratio of the vacuum expectation values  (VEVs) of the two MSSM Higgs
doublets, and the magnitude of $\mu$, but not its sign, is determined by the
radiative electroweak breaking (REWSB) condition.

 Whereas, universal scalar masses are motivated to suppress unwanted flavor changing neutral currents, the Higgs mass parameters can be non-universal. The effects of this non-universality on the parameter space has been studied in models called Non-Universal Higgs Models (NUHM) \cite{Baer:2008ih}. One of the types of these models called the NUHM2 has two additional parameters compared to the CMSSM
\begin{equation}
m_{H_u}^2, m_{H_d}^2,
\label{nuhm2-params}
\end{equation}
where $m_{H_u}^2$ and $m_{H_d}^2$ are  the SSB the MSSM  Higgs mass$^2$ term.
 The supersymmetric flipped SU(5) (FSU(5)) model \cite{DeRujula:1980qc} is based on the maximal
subgroup $G\equiv {\rm SU(5)}\times {\rm U(1)}_X$ of SO(10), and the sixteen
chiral superfields per family of SO(10) are arranged under $G$ as:
$10_1 =  ( d^c,  Q, \nu^c)$, $\bar{5}_{-3} = (u^c , L)$,
$1_5= e^c$. Here the subscripts refer to the respective charges under
${\rm U(1)}_X$, and we follow the usual notation for
the Standard Model (SM) particle
content. The MSSM electroweak Higgs doublets $H_u$ and $H_d$ belong
to $\bar 5_H$ and $5_H$ of SU(5), respectively. We will assume for
simplicity that the soft mass$^2$ terms, induced at $M_{\rm GUT}$ through
gravity mediated supersymmetry breaking \cite{Chamseddine:1982jx},
are equal in magnitude for the scalar squarks and sleptons of the
three families. The asymptotic MSSM gaugino masses, on the other
hand, can be non-universal. Due to the FSU(5) gauge structure,
 asymptotic ${\rm SU(3)}_c$ and ${\rm SU(2)}_W$ gaugino masses can be
different from the ${\rm U(1)}_Y$ gaugino mass. Assuming SO(10) normalization for
${\rm U(1)}_X$, the hypercharge generator in FSU(5) is given
by $Y=(-{Y_5}/2+ \sqrt{24} X)/5$, where $Y_5$
and $X$ are the  generators of SU(5) and ${\rm U(1)}_X$ \cite{Barr:2006im}. We
then have the following asymptotic relation between the three MSSM gaugino masses:
\begin{align}
M_1=\frac{1}{25} M_5 + \frac{24}{25} M^\prime,\; \mbox{with} \;
M_5=M_2=M_3, \label{gauginoCondition}
\end{align}
where $M_5$, $M^\prime$, $M_3$, $M_2$ and $M_1$ denote SU(5),  ${\rm U(1)}_X$, $SU(3)_c$, $SU(2)_L$ and $U(1)_Y$ gaugino
masses respectively. The supersymmetric FSU(5) model  thus has two
independent parameters $(M_2=M_3,\, M^\prime)$ in the gaugino
sector. In other words, in FSU(5), by assuming gaugino
non-universality, we increase by one the number of fundamental
parameters compared to the CMSSM

We will also consider both universal (${m^2_{H_u}}$$=$${m^2_{H_d}}$)
and non-universal (${m^2_{H_u}}$$\neq$${m^2_{H_d}}$) soft scalar
Higgs masses in FSU(5), using the notations for FSU(5)-UH and FSU(5) respectively.   This  would mean up to three additional parameters compared to
the CMSSM. This latter case, with one additional gaugino mass parameter and two soft
scalar mass parameters, provides us with a compelling neutralino
dark matter candidate for indirect and direct detection \cite{Gogoladze:2009mc} in the ongoing and future
experiments.

We use the ($\mu,m_A$) parameterization to characterize non-universal
soft scalar Higgs masses rather than ($H_u,H_d$). The fundamental parameters of our
FSU(5) model are
\begin{align}
m_0,M^\prime,M_2,{\rm tan\beta},A_0,\mu,m_A, \label{params}
\end{align}
we will assume that $\mu>0$. Note that $\mu$ and
$m_A$ are specified at the weak scale, whereas the other parameters are
specified at $M_{\rm GUT}$. Although not required, we will assume that
the gauge coupling unification condition $g_3=g_1=g_2$ holds at $M_{\rm GUT}$
in FSU(5). Such a scenario can arise,
for example, from a higher dimensional
theory \cite{Jiang:2009za} after suitable choice of
compactification.

\section{Muon Flux, SD cross section and Conversion factors}\label{flux-sd}
In this section we will review the calculation of the muon flux and spin dependent cross section and revisit the way the conversion factors between the two are calculated. The IceCube collaboration \cite{Abbasi:2009uz} has presented its results as a future bound on the muon flux from the Sun. The bound was then converted to a bound on the spin dependent cross section by suitable conversion factors calculated in \cite{Wikstrom:2009kw} and also discussed in \cite{Wikstrom:2009zz}.

The flux of neutrino induced muons from neutralino annihilation in the Sun is given by
\begin{eqnarray}
 \Phi_{\mu}=\frac{\Gamma_{A}\cdot n}{4\pi D_{\odot}^{2}}\int_{E_{\mu}^{th}}^{\infty}\mathrm{d}E_{\mu}\int_{E_{\mu}}^{\infty}\mathrm{d}E_{\nu} \int_{0}^{\infty}\mathrm{d}\lambda
\int_{E_{\mu}}^{E_{\nu}}  \mathrm{d}E_{\mu}^{'} P_{\textrm{\tiny SURV}}(E_{\mu} ,E_{\mu}^{'},\lambda)\times\frac{\mathrm{d}\sigma_{\nu}(E_{\nu},E_{\mu}^{'})}{\mathrm{d}E_{\mu}^{'}}
 \sum_{i}P{\textrm{\tiny osc}}(\mu,i)\sum_{f}B_{f}\frac{\mathrm{d}N_{i}^{f}}{\mathrm{d}E_{\nu}}.
\label{eq:muflux}
\end{eqnarray}
Here $\Gamma_{A}$ is the annihilation rate, {\it n} is the target number density, and $D_{\odot}$ is the distance from the center of the Sun to the detector. $dN^{f}_{i}/dE_{\nu}$ is the differential energy spectrum of the number of neutrinos from neutralino annihilation  with the corresponding branching fractions $B_f$. $E_{\mu}^{th}$ is the threshold energy of the muon in the detector, $\lambda$ is the muon range, $P_{\textrm{\tiny SURV}}(E_{\mu},E_{\mu}^{'},\lambda)$ is the survival probability for a muon, $\mathrm{d}\sigma_{\nu}(E_{\nu},E_{\mu}^{'})/\mathrm{d}E_{\mu}^{'}$ is the differential neutrino cross-section and $P(\mu,i)$ is the oscillation probability for a neutrino of flavor $i$ to oscillate to flavor $\mu$ in the detector.

The annihilation rate for the weakly interacting massive particles  (WIMPs) in the center of the Sun is given by
\begin{equation}
\Gamma_{A}=\frac{1}{2} C_C \rm tanh^{2}(t/\tau),
\end{equation}
$\tau=(C_{C}C_{A})^{-1/2}$ is a measure of the time in which capture and annihilation equilibrate, $C_{C}$ is the capture rate, and $C_A$ parameterizes the annihilation rate of the WIMPs. For present WIMP annihilation rate, $t$ is the age of the Sun, i.e., $t=t^{\odot}\simeq4.5\cdot10^{9}$ years. The annihilation                                                                                                                                                                                                                                                                                                                                                                                                                                                      and capture rate are in equilibrium when $t^{\odot}/\tau \gg1$, which implies
\begin{equation}
 \Gamma_{A}=\frac{1}{2}C_{C}.
\end{equation}
Since the capture rate is proportional to the spin dependent and spin independent cross sections, there would be a direct correlation between the flux and the SD cross section. The converted bound cannot be trusted for models where the equilibrium condition is not satisfied.

Accurate expressions for the capture rate can be found in \cite{Gould:1987ir}, while reference \cite{Jungman:1995df} gives the approximate expressions. For the case of SD cross section, which occurs mainly on hydrogen and the form factor suppression is negligible, the capture rate in the Sun can be written as
\begin{eqnarray}
 C^{\odot}_{SD}=(1.3\cdot 10^{23}\, {\rm s}^{-1})\left(270\ {\rm km\,s^{-1}}\over \bar v\right) \left({\rho_\chi\over 0.3\ {\rm GeV}\,{\rm cm}^{-3}}\right)
 \left({100\,{\rm GeV}\over m_\chi}\right)\left({\sigma_{\rm SD}\over 10^{-40}\ {\rm
cm}^2}\right) S(m_\chi /m_p),
\label{caprate}
\end{eqnarray}
where $\sigma_{\rm SD}$ is the neutralino-proton spin dependent cross section, $\bar v=270 {\rm \;Km \,s^{-1}}$ is the dark matter velocity dispersion, $\rho_{\chi}=0.3{\rm \;GeV \,cm^{-3}}$ is the local dark matter density, and $S(m_\chi /m_p)$ is the kinematical suppression factor defined as
\begin{equation}
S(x)=\left(\frac{A^{3/2}}{1+A^{3/2}}\right)^{2/3},
\end{equation}
with
\begin{equation}
A(x)=\frac{3x}{(x-1)^2}\left(\frac{\langle v^2_{esc} \rangle}{\bar v^2}\right).
\end{equation}
$\langle v_{esc} \rangle$ denotes the mean escape velocity from the Sun.

\begin{figure}[!ht]
\includegraphics[scale=0.9]{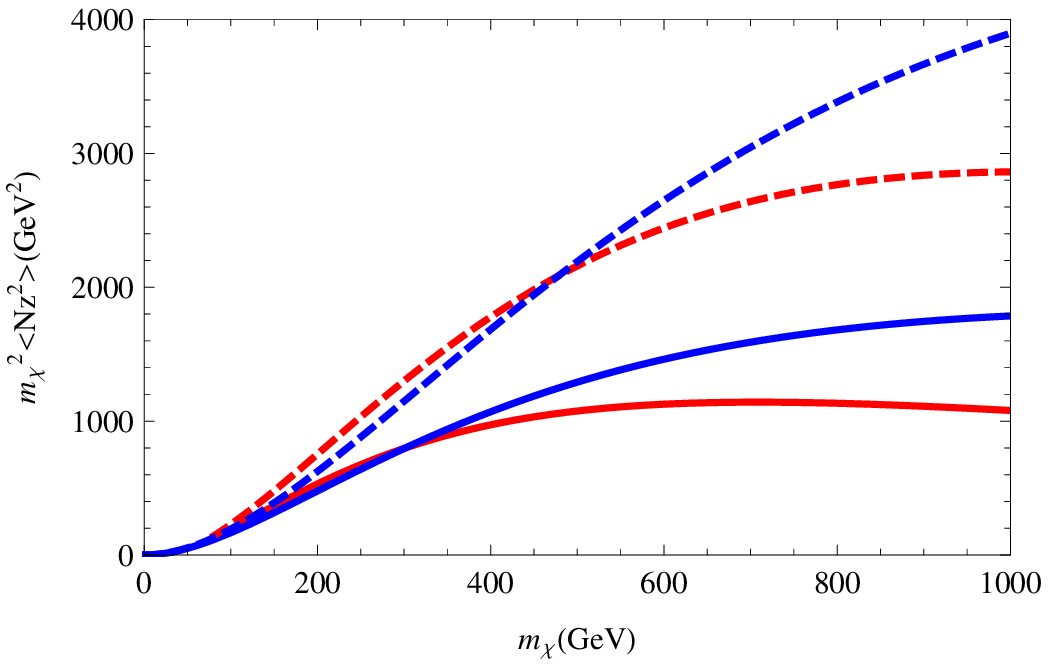}
\caption{
Second moment of the neutrino spectrum. The solid lines correspond to neutrinos
and the dashed lines are for anti-neutrinos. The $W^{+}W^{-}$ channel is in red and
$\tau^{+}\tau^{-}$ channel is in blue.
\label{fig:6}}
\end{figure}

Reference \cite{Wikstrom:2009kw} calculates accurate conversion factors including neutrino oscillations. Here we take a simple example to see how the conversion factors are calculated to get the SD cross section from the muon flux. Ignoring detector thresholds and taking the effective range of muons in the detector, the rate of neutrino induced through going muons for the Sun can be approximated as \cite{Jungman:1995df}
\begin{eqnarray}
\Gamma_{\mu}\approx(1.27\times 10^{-23} {\rm km^{-2} yr^{-1}})\frac{C_C}{s^{-1}}\left(\frac{m_{\chi}}{1 {\rm GeV}}\right)^2
\sum_{i}a_{i}b_{i}\sum_{F}B_f\langle Nz^2 \rangle_{f,i}(m_{\chi}).
\label{muflux2}
\end{eqnarray}
Since the capture in the Sun is mainly through spin dependent scattering, we can assume $\sigma_{SI}=0$ ($C_C=C^{\odot}_{SD}$) to get a bound
on the SD cross section. $a_{i}$ are the neutrino scattering coefficients $a_{\nu}=6.8$ and $a_{\bar {\nu}}=3.1$, and $b_{i}$ are the muon range coefficients with $b_{\nu}=0.51$ and $b_{\bar {\nu}}=0.67$. The quantity $\langle Nz^2 \rangle_{f,i}$ is the second moment of the neutrino spectrum of type $i$ from final state $f$, scaled by the square of the injection
energy $E_{in}$ of the annihilation products, and is given by
\begin{equation}
\langle Nz^2 \rangle_{f,i}(m_{\chi})=\frac{1}{E^2_{in}}\int \left(\frac{dN}{dE}\right)_{f,i}
(E_{\nu},E_{in})E^2_{\nu} dE_{\nu}.
\label{nz2}
\end{equation}
The neutrino spectrum from the $W^{+}W^{-}$ and $\tau^{+}\tau^{-}$ channels can be taken as  \cite{Jungman:1994jr}
\begin{eqnarray}
\left(\frac{dN}{dE_{\nu}}\right)^{\odot}_{WW}=\frac{\Gamma_{W\rightarrow \mu \nu}}{E_{in}}
(1+E_{\nu}\tau_{i})^{-\alpha_{i}-2},
\label{eq:wwspect}
\end{eqnarray}
with $E_{in}(1-\beta/2) \leq E_{\nu} \leq E_{in}(1+\beta/2)$, $\Gamma_{W\rightarrow \mu \nu}=0.105$
, $\beta=(1-m^2_W/E^2_{in})^{1/2}$, $\tau_{\nu}=1.01\times 10^{-4} \rm{GeV^{-1}}$ and
$\tau_{\bar{\nu}}=3.8\times 10^{-4} \rm{GeV^{-1}}$, and
\begin{eqnarray}
\left(\frac{dN}{dE_{\nu}}\right)^{\odot}_{\tau^{+}\tau^{-}}=\frac{2\Gamma_{\tau\rightarrow \mu\nu\nu}}{E_{in}}
(1-3x^2+2x^3) (1+E_{\nu}\tau_{i})^{-\alpha_{i}-2},
\label{eq:tauspect}
\end{eqnarray}
with $0\leq E_{\nu}\le E_{in}$ $x=E_{\nu}/E_{in}$ and $\Gamma_{\tau\rightarrow \mu\nu\nu}=0.18$, $\alpha_{\nu}=5.1$ and $\alpha_{\bar{\nu}}=9.0$.
Note that improved functions for the spectra can be obtained by using programs like Pythia. In Fig.\ref{fig:6} we show plots of the second moment of these functions. We take $B_f=1$ and only consider contributions of the hard channels $W^{+}W^{-}$ and $\tau^{+}\tau^{-}$. In our plots we assume that only the  $W^{+}W^{-}$ channel contributes for $m_{\chi}>80 {\rm {\, GeV}}$, and the $\tau^{+}\tau^{-}$ channel for $m_{\chi}<80 {\rm {\, GeV}}$.  The second moments $\langle Nz^2 \rangle_{WW}$ and $\langle Nz^2 \rangle_{\tau\tau}$ for the $W^{+}W^{-}$ and $\tau^{+}\tau^{-}$ channels are obtained by inserting Eqs. (\ref{eq:wwspect}) and (\ref{eq:tauspect}) in (\ref{nz2}). From Eq. (\ref{muflux2}),
\begin{eqnarray}
\Gamma_{\mu}=(1.27\times 10^{-23} {\rm km^{-2} yr^{-1}})\frac{ C^{\odot}_{SD}}{s^{-1}}\left(\frac{m_{\chi}}{1 {\rm GeV}}\right)^2
[3.47 \langle Nz^2 \rangle_{WW,\nu}(m_{\chi})+2.08\langle Nz^2 \rangle_{WW,\bar{\nu}}(m_{\chi)}],
\label{muflux3}
\end{eqnarray}
for $m_{\chi}>(<)80{\rm {\, GeV}}$ for the $W^{+}W^{-}$ ($\tau^{+}\tau^{-}$) channel.

\begin{figure*}[!h]
\includegraphics[scale=1.3]{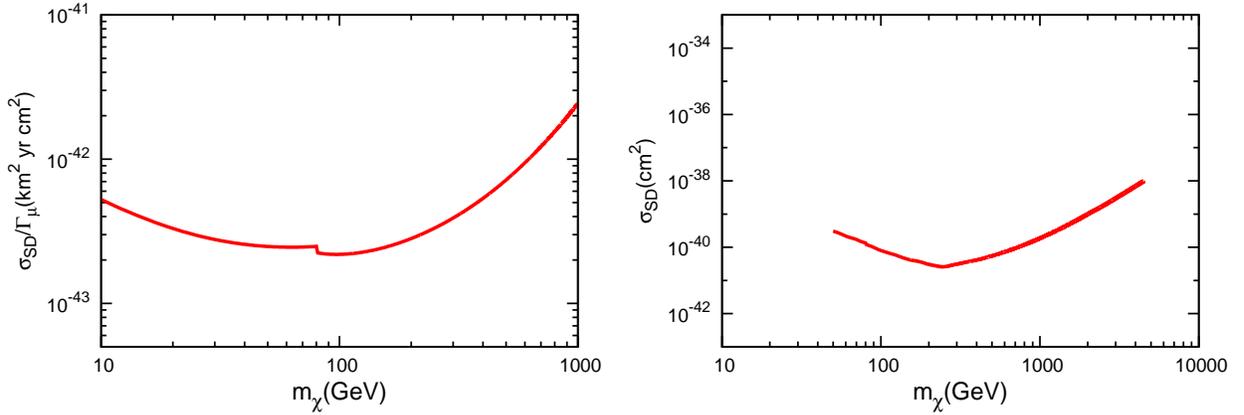}
\caption{
The left panel shows the conversion factor calculated from approximate expressions for the
flux and spin dependent cross section. The right panel shows the converted future IceCube/DeepCore
muon flux bound using this conversion factor.
\label{fig:1}}
\end{figure*}

 We can re-write Eq. (\ref{caprate}) as
\begin{equation}
 C^{\odot}_{SD}=f_1(m_\chi) \sigma_{\rm SD},
\end{equation}
and inserting this in Eq. (\ref{muflux2}) we have
\begin{equation}
\Gamma_{\mu}=f_1(m_\chi)f_2(m_\chi) \sigma_{\rm SD},
\end{equation}
which yields the conversion factor
\begin{equation}
\frac{\sigma_{\rm SD}}{\Gamma_{\mu}}=f^{-1}_1(m_\chi)f^{-1}_2(m_\chi) (\rm{km^2\;yr\;cm^{2}}).
\end{equation}
The conversion factor and the converted future IceCube/DeepCore bound obtained from it are plotted in Fig. \ref{fig:1}. The conversion factor we find is similar to the estimates made in \cite{Wikstrom:2009kw} without including neutrino oscillations. We will see later how the converted IceCube/DeepCore future bound for the SD cross section is changed once  neutrino oscillations are included.

Neutrinos created from neutralino annihilations in the Sun's center may experience oscillations and interactions (neutral as well as charged) in the Sun. These and other factors, like the Solar composition, can modify the flux observed by detectors on Earth. On the other hand, direct detection involves measuring the scattering cross section of neutralinos off of nuclei. The conversion of the limit obtained from indirect detection experiments thus involves several uncertainties  \cite{Wikstrom:2009kw}. They include:
\begin{itemize}
\item Uncertainties in the Solar model.
\item Gravitational effects of planets like Jupiter.
\item Form factors.
\item Variations in local dark matter density and velocity distributions.
\item Neutrino oscillations.
\end{itemize}
These uncertainties can affect the estimated muon flux and thereby the deduced cross section. Gravitational affects from Jupiter, for example, can reduce the estimated muon flux, whereas neutrino oscillations can enhance it. The form factor suppression is negligible for the case of spin dependent interaction since the capture in the Sun mainly occurs through scattering on the hydrogen nuclei. All of the above listed affects can have implications for the particular particle physics model being investigated. The implications for our models are discussed in section \ref{sec:results}.
\section{Phenomenological constraints and scanning procedure}\label{procedure}
We use the package DarkSUSY-5.0.5 \cite{Gondolo:2004sc} to calculate the flux of neutrino induced muons from the Sun. DarkSusy (DS) uses a local dark matter density of $\rho_{\chi}=0.3 \textrm{ GeV/cm}^{3}$. From the three methods DS employs to calculate the relic density, we pick the one which includes coannihilations only if the mass difference between the LSP and NLSP is less than $30\%$.  We rescale the neutralino density to $\rho=\rho_0 (\Omega h^2/0.025)$, for $\Omega h^2<0.025$. This is done because the neutralino cannot make up all of the dark matter in the galaxy halos if  $\Omega h^2$ drops below 0.025.

  Although not required, we assume for simplicity, that the gauge coupling unification condition $g_1=g_2=g_3$ holds at $M_{GUT}$ for FSU(5). DarkSUSY-5.0.5 uses Isajet 7.78 \cite{Baer:1999sp} for renormalization group evolution (RGE) running and the latter employs two loop MSSM RGEs and defines $ M_{GUT} $ to be the scale where $ g_{1}=g_{2} $. A few percent deviation from exact unification ($g_1=g_2=g_3$) can be attributed to unknown GUT-scale threshold corrections~\cite{Hisano:1992jj}.

 For the random scan we employ the following ranges for our parameters
\begin{eqnarray}
0\leq  m_{0} \leq 5\, \rm{TeV}, \nonumber \\
0\leq  M^\prime \leq 1\, \rm{TeV}, \nonumber  \\
0\leq  M_{2} \leq 1\, \rm{TeV}, \nonumber  \\
0 \leq m_{A} \leq 1\, \rm{TeV}, \nonumber \\
0 \leq \mu  \leq 10\, \rm{TeV}, \nonumber \\
{\rm tan\beta}=10,30,50, \nonumber \\
A_{0}=0,
\label{parameterRange}
\end{eqnarray}
we set $ m_{t}=173.1 \textrm{ GeV} $.

The random scan is performed over the parameter space of CMSSM ($M_2$=$M_3$=$M'$=$m_{1/2}$, $m_0$=$m_{H_u}$=$m_{H_d}$), FSU(5), NUHM2
($M_2$=$M'$), and FSU(5) with universal soft Higgs $\rm {masses}^2$ ($m^2_{H_u}$=$m^2_{H_d}$) at $M_{GUT}$. Since the neutralino mass is sensitive to the gaugino masses, we manipulate the latter to obtain more allowed points in the parameter space. We take piece-wise intervals [0,10],[10,100] and [100,1000] for the gaugino masses (in units of GeV). The random points in each of these intervals are distributed logarithmically. These points were then combined with a uniform distribution of points on the interval [0,1000], with the total number of points around a million, which enables us to obtain a sufficiently dense set of points for our plots. This is
still not sufficient for FSU(5)-UH, so we perform a Gaussian scan around the allowed points. The code makes a Gaussian distribution of points for the scalar and gaugino mass parameters around a point satisfying all the imposed constraints, with the variance and mean of the Gaussian distribution being 1/25 and 1 respectively. The random function RNORMX(), available in the program library of CERN, was used to make this Gaussian distribution.

We apply the experimental constraints on the data sequentially, with all of the collected data points satisfying the requirement of radiative electroweak symmetry breaking (REWSB), and the neutralino in each cases being the LSP.
On this data, we impose the   following constraints:
\begin{table}[h!]\centering
\begin{tabular}{rlc}
$m_{\tilde{\chi}^{\pm}_{1}}~{\rm (chargino~mass)}$ & $ \geq\, 103.5~{\rm GeV}$ &  \cite{Yao:2006px}      \\
$m_{\tilde \tau}~{\rm (stau~mass)} $&$ \geq\, 105~{\rm GeV}$                     &   \cite{Yao:2006px}   \\
$m_{\tilde g}~{\rm (gluino~mass)} $&$ \geq\, 250~{\rm GeV}$                     &    \cite{Yao:2006px}  \\
$m_{\tilde t}~{\rm (stop~mass)} $&$ \geq\, 175~{\rm GeV}$                     &   \cite{Yao:2006px}   \\
$m_{\tilde b}~{\rm (sbottom~mass)} $&$ \geq\, 222~{\rm GeV}$                     &   \cite{Yao:2006px}   \\
$m_h~{\rm (lightest~Higgs~mass)} $&$ \geq\, 114.4~{\rm GeV}$                    &  \cite{Schael:2006cr} \\
$BR(B_s \rightarrow \mu^+ \mu^-) $&$ <\, 5.8 \times 10^{-8}$                     &   \cite{:2007kv}      \\
$2.85 \times 10^{-4} \leq BR(b \rightarrow s \gamma) $&$ \leq\, 4.24 \times 10^{-4} \; (2\sigma)$ &   \cite{Barberio:2007cr}  \\
$\Omega_{\rm CDM}h^2 $&$ =\, 0.111^{+0.028}_{-0.037} \;(5\sigma)$               &  \cite{Komatsu:2008hk}    \\
\end{tabular}
\end{table}

Note that we do not include the $(g-2)_{\mu }$ constraint for the rest of our discussion.
\section{Results}\label{sec:results}

\begin{figure*}[!h]
\includegraphics*[scale=0.7]{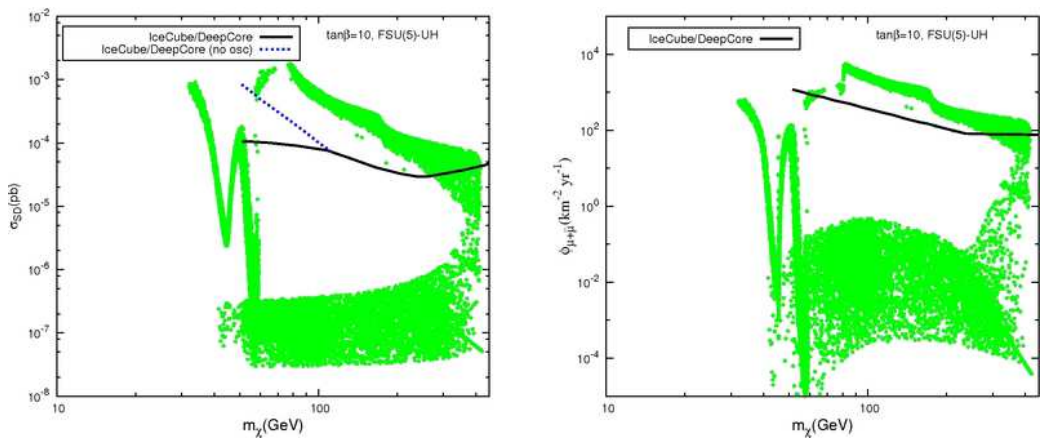}
\caption{
Comparison of the future IceCube/DeepCore muon flux bound and the converted SD cross section bound. The muon flux is from the Sun above 1 GeV threshold is shown for ${\rm tan\beta}=10$. The dashed line in the left panel is the future IceCube/DeepCore bound obtained if neutrino oscillations are not
included in the flux calculation. The conversion factors used are given in reference \cite{Wikstrom:2009kw}.
\label{fig:2}}
\end{figure*}

We next present the results from the scan over the parameter space listed in  Eq. (\ref{parameterRange}). In Fig.~\ref{fig:2}, we show  how the converted IceCube/DeepCore future bound is altered with the inclusion of neutrino oscillations.
The colored points
 are consistent with REWSB and satisfy the WMAP relic density bound in the 5$\sigma$ range, particle mass bounds, and all constraints coming from the B-physics. We used different color coding to distinguish different channels for neutralino dark matter.
We see from the right panel in  Fig.~\ref{fig:2}, and as noted in reference \cite{Wikstrom:2009kw}, the inclusion of neutrino oscillations has a notable affect on the bound, especially for neutralino mass less than the W boson mass. As the W boson decouples ($m_{\chi}<80{\rm \,GeV}$)  and the contribution from the tau channel becomes relevant, the bound changes notably. The reason for this is that the tau neutrinos from the decay $\tau^{-}\rightarrow \mu^{-}\bar {\nu}_{\mu}\nu_{\tau}$ can oscillate into muon neutrinos, thereby enhancing the muon flux in the detector. This is especially relevant for the low neutralino masses we have in FSU(5).
In the left  panel the future IceCube/DeepCore bound (solid line) indicates that the light neutralino ($m_{\chi}< 70$ GeV) parameter space can be tested at IceCube/DeepCore detector, but this same region of the parameter space, it seems, does not yield sufficient muon fluxes, as can be seen  in the right panel of Fig.~\ref{fig:2}. So we see that there is some discrepancy.

Note that, whereas the calculation of the flux is sensitive to the various channels from which the neutrino arises, the cross section is not. Fig.~\ref{fig:2} shows that as the neutralino mass falls below the W mass, and only soft channels (e.g. $b\overline{b}$) are left , the muon flux starts decreasing, whereas the SD cross section decreases much less rapidly \cite{private}. This is because the SD cross section is not prone to the hardness or softness of the channel, which is not the case for the bound on the SD cross section where the sensitivity of the flux is translated to the SD cross section.

\begin{figure*}[!h]
\includegraphics*[scale=0.6]{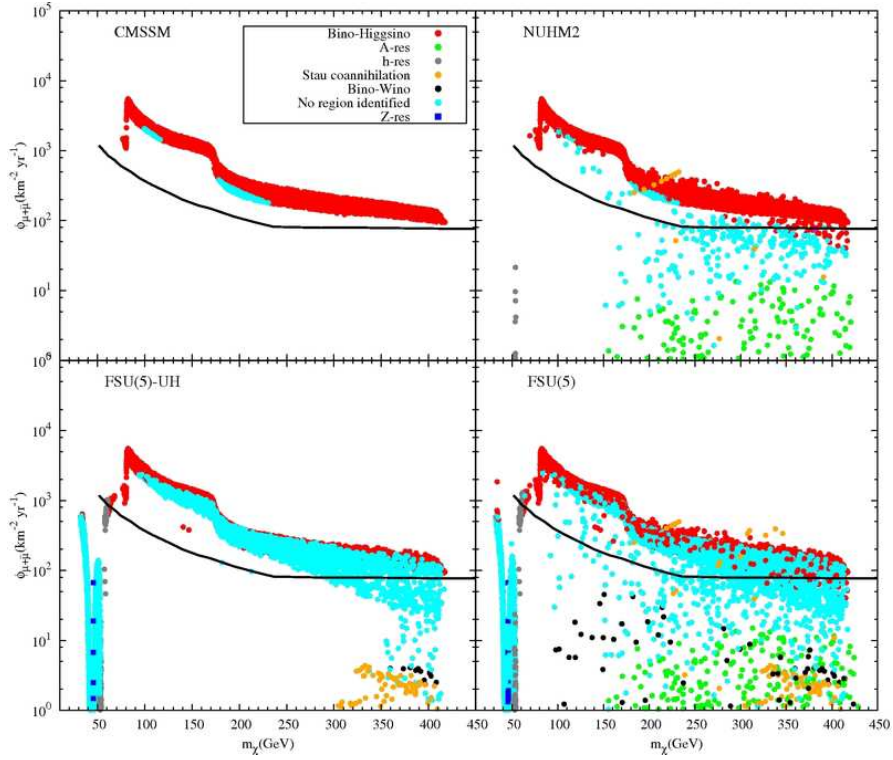}
\caption{
Flux of $\mu+\overline{\mu}$ from the Sun above 1 GeV threshold for ${\rm tan\beta}=10$. The black line shows the future IceCube/DeepCore bound \cite{Abbasi:2009uz}. The  colored points  are consistent with REWSB and satisfy the WMAP relic density  bound in the 5 $\sigma$ range, particle mass bound, and all constraints coming from B-physics. The points in different colors correspond to the various solutions of LSP  neutralino to be a dark matter  candidate.
\label{fig:3}}
\end{figure*}

\begin{figure*}[tp]
\includegraphics*[scale=0.6]{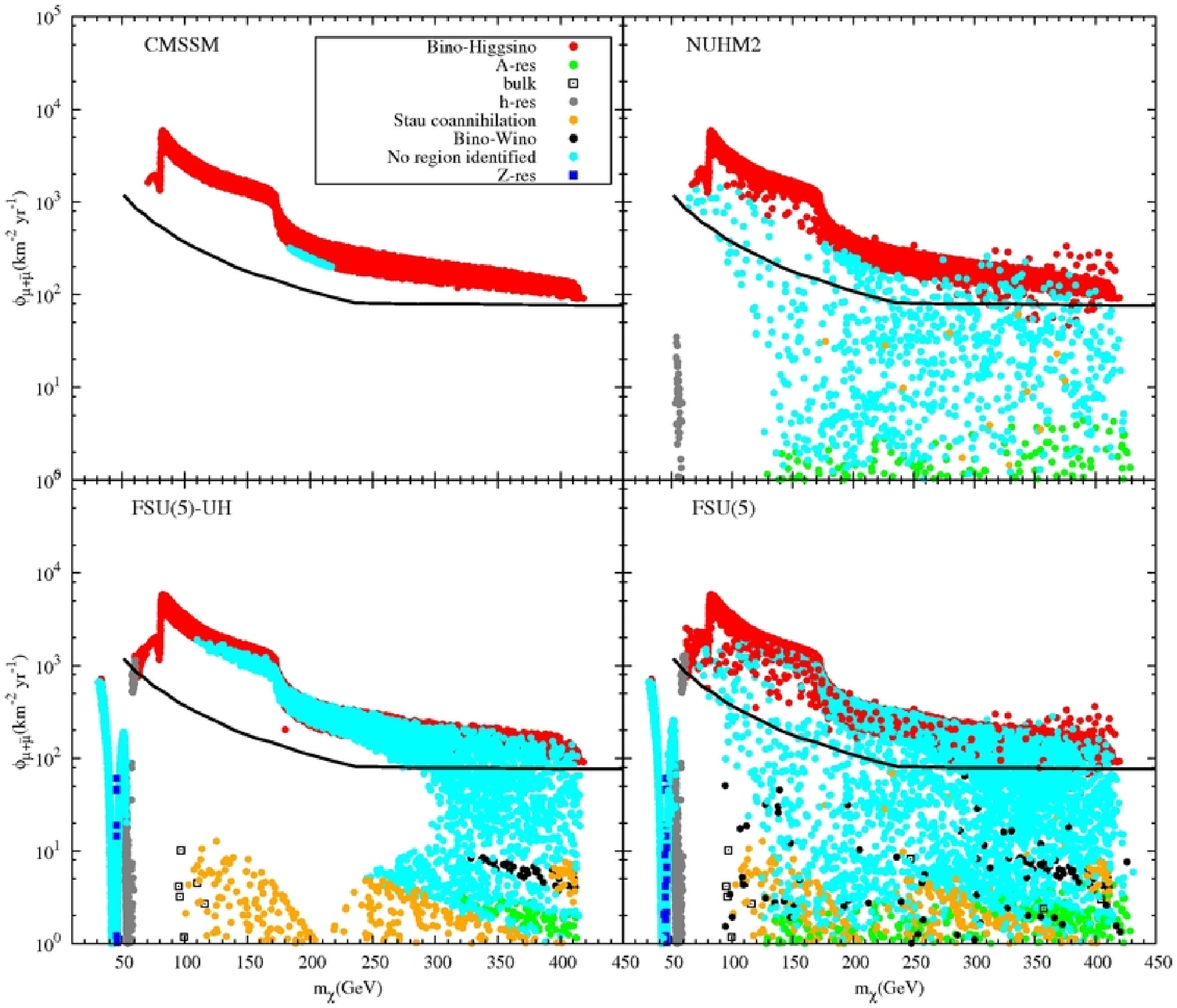}
\caption{
Flux of $\mu+\overline{\mu}$ from the Sun above 1 GeV threshold for ${\rm tan\beta}=30$. We use the same color coding as in Fig. \ref{fig:3}.
\label{fig:4}}
\end{figure*}

\begin{figure*}[h]
\includegraphics*[scale=0.6]{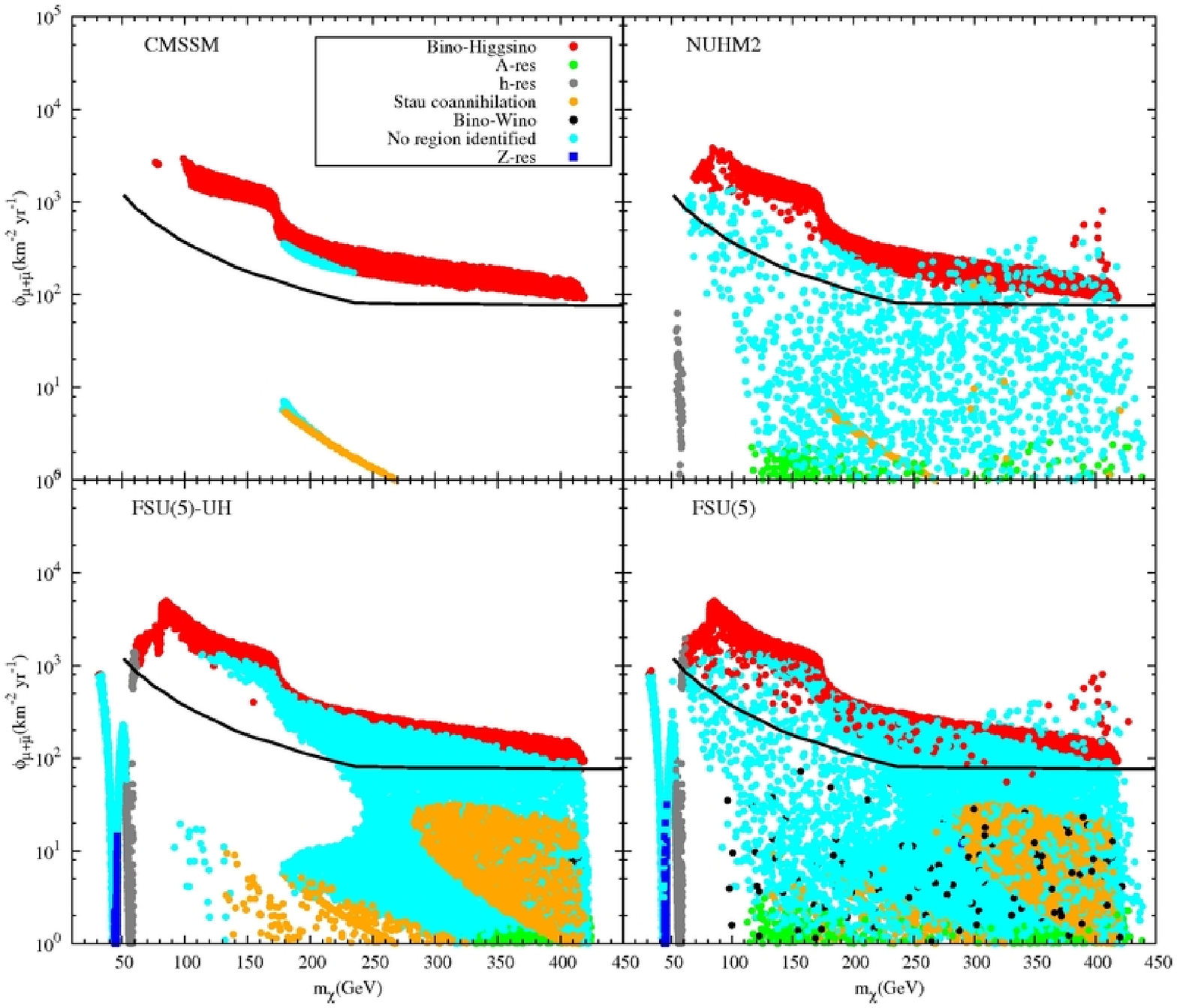}
\caption{
Flux of $\mu+\overline{\mu}$ from the Sun above 1 GeV threshold for ${\rm tan\beta}=50$. We use the same color coding as in Fig. \ref{fig:3}.
\label{fig:5}}
\end{figure*}

For the calculation of the flux we use the 'default' method in DS which uses the approximate expression for the capture rate in the Sun  from \cite{Jungman:1995df}.
It is understood that the dark matter prediction is no longer a natural consequence of supersymmetry \cite{ArkaniHamed:2006mb}, but requires special relations among the parameters. To have the correct relic dark matter abundance, we require coannihilation, resonance  or specific Bino-Higgsino mixing solution.  On the other hand this then yields some very specific structure for the sparticle spectroscopy which can be tested at the LHC. This explains why, in Figs.  \ref{fig:3}, \ref{fig:4} and \ref{fig:5}, we show the various relic channels for neutralino dark matter. The colored points
 are consistent with REWSB and satisfy the WMAP relic density  bound in the 5 $\sigma$ range, particle mass bounds, and all constraints from B-physics. Figs. \ref{fig:3}, \ref{fig:4} and \ref{fig:5} present the muon flux induced by the neutrinos originating from annihilating neutralino dark matter in the center of the Sun, for ${\rm tan\beta}=10,\,30,\,50$. The points shown satisfy the WMAP relic density bounds in the 5$\sigma$ range. The calculated muon flux is integrated above a threshold energy of 1 GeV. From Figs. \ref{fig:3}, \ref{fig:4} and \ref{fig:5}, we observe that the IceCube/DeepCore detector  can test the following neutralino dark matter scenarios: Bino-Higgsino dark matter, light Higgs resonance and finally the "non-identified channel", which is a combination of various channels. The points we designate as being non-identified means that the conditions we apply on the neutralino to be from all other channels are not satisfied. Thus an observed signal at the IceCube/DeepCore detector can narrow the probable  neutralino dark matter candidates, and combining this with a signal from the LHC may help identify the nature of dark matter.

\begin{figure*}[t]
\includegraphics*[scale=0.7]{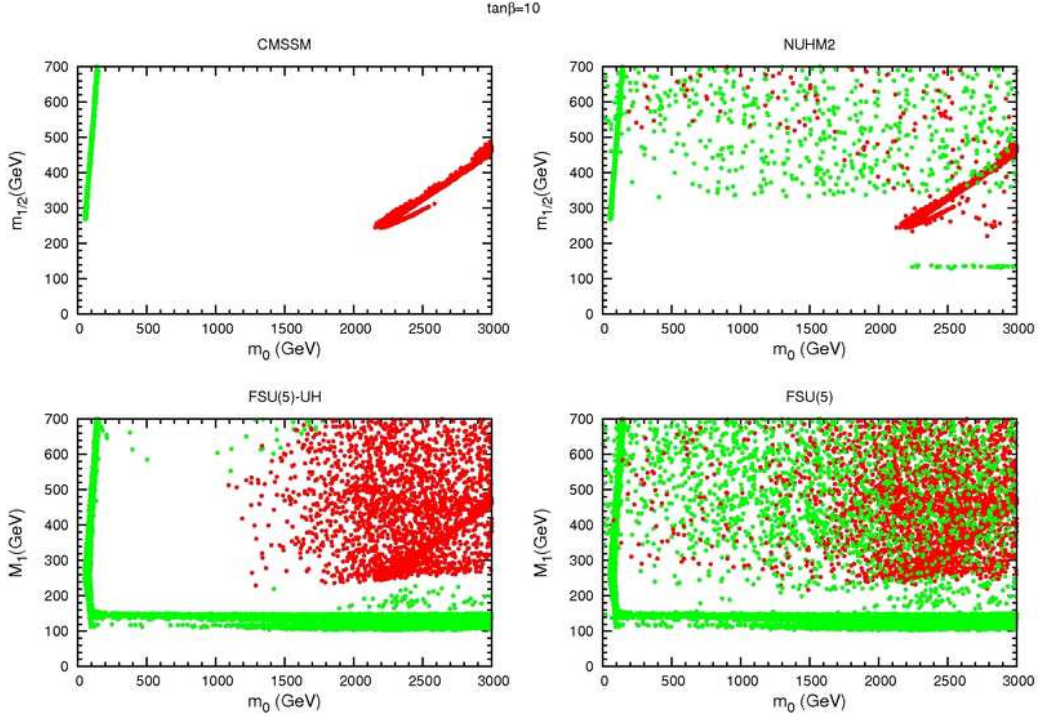}
\caption{Plot in the  $m_{1/2}$ and $M_1$ vs $m_{0}$ plane. We are comparing the allowed parameter spaces for differed models.
The  green points   are consistent with REWSB and satisfy the WMAP relic density  bound in the 5 $\sigma$ range, particle mass bound, and all constraints coming from B-physics.
 The red points are a subset of the green ones and can generate detectable muon fluxes at the IceCube/DeepCore detector.
\label{fig:7}}
\end{figure*}

In Fig. \ref{fig:7} we show the results in  the fundamental parameter planes.  Here $M_{1/2}$ stands for the GUT scale universal gaugino mass in CMSSM and NUHM2, and $M_1$ is the Bino mass for FSU(5)-UH  and FSU(5) plots, the expression for which is given in  Eq.~(\ref{gauginoCondition}).
The green points are consistent with REWSB and satisfy the WMAP relic density  bound in the 5 $\sigma$ range, particle mass bound, and all constraints coming from B-physics. The red points are a subset of the green ones and give muon fluxes which can be tested at the IceCube/DeepCore experiment.  The IceCube/DeepCore experiment, we see, can test a significant region of the flipped SU(5) parameter space.

\begin{figure*}[t]
\includegraphics*[scale=0.7]{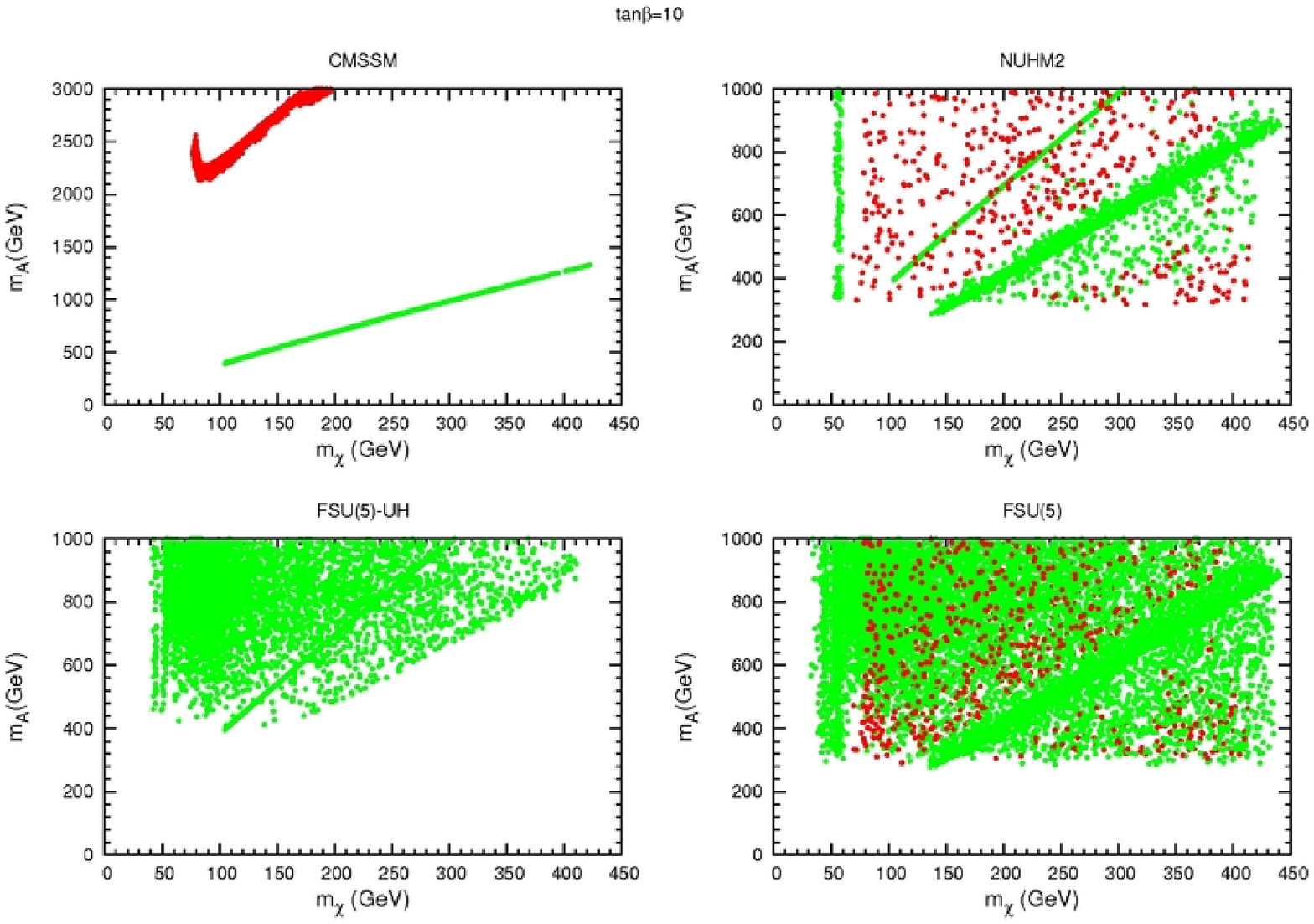}
\caption{Plot in the $m_A - m_{\chi}$ plane. Color coding same as in Fig. \ref{fig:7}
\label{fig:8}}
\end{figure*}

\begin{figure*}[t]
\includegraphics*[scale=0.7]{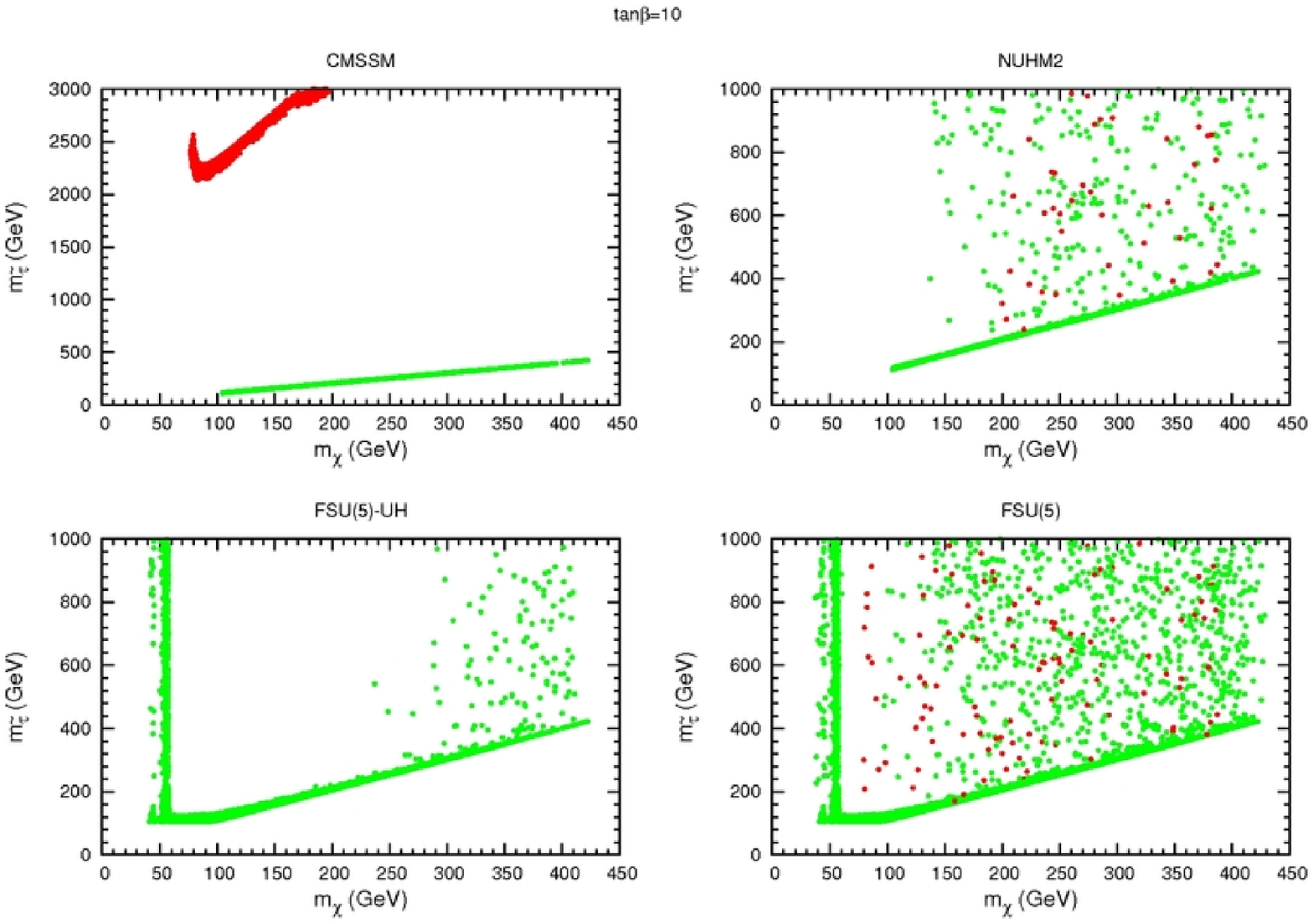}
\caption{Plot in the $m_{\tilde{\tau}} - m_{\chi}$ plane. Color coding same as in Fig. \ref{fig:7}
\label{fig:9}}
\end{figure*}

\begin{figure*}[t]
\includegraphics*[scale=0.7]{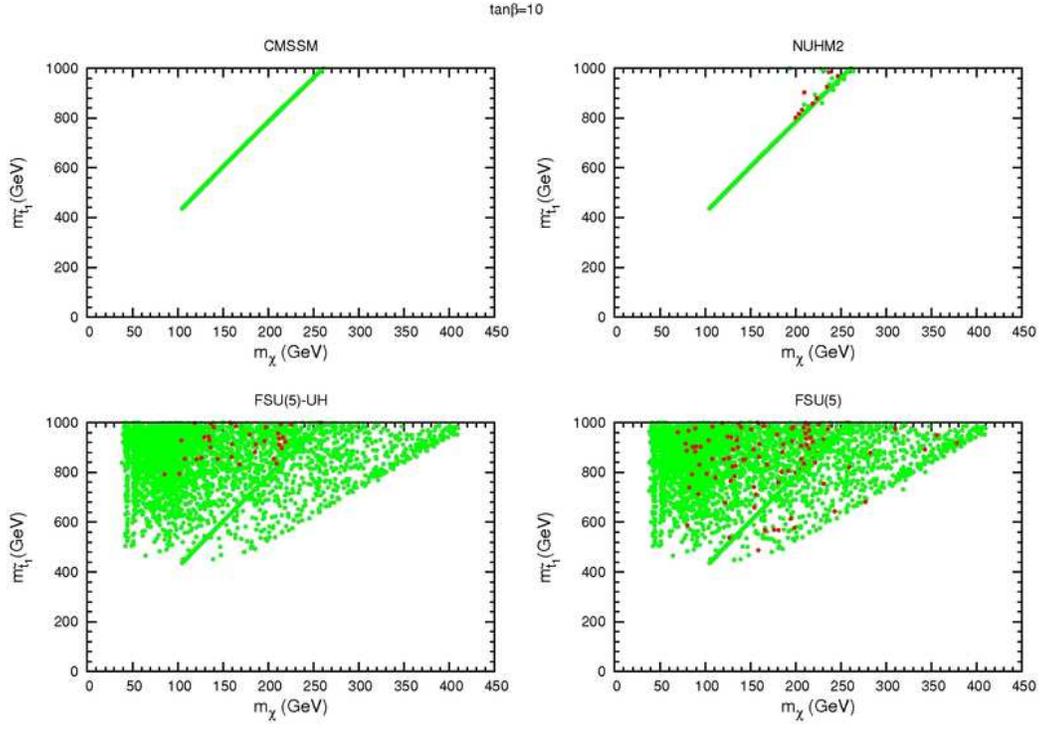}
\caption{Plot in the  $m_{\tilde{t}_1} - m_{\chi}$ plane. Color coding same as in Fig. \ref{fig:7}
\label{fig:10}}
\end{figure*}

\begin{figure*}[t]
\includegraphics*[scale=0.7]{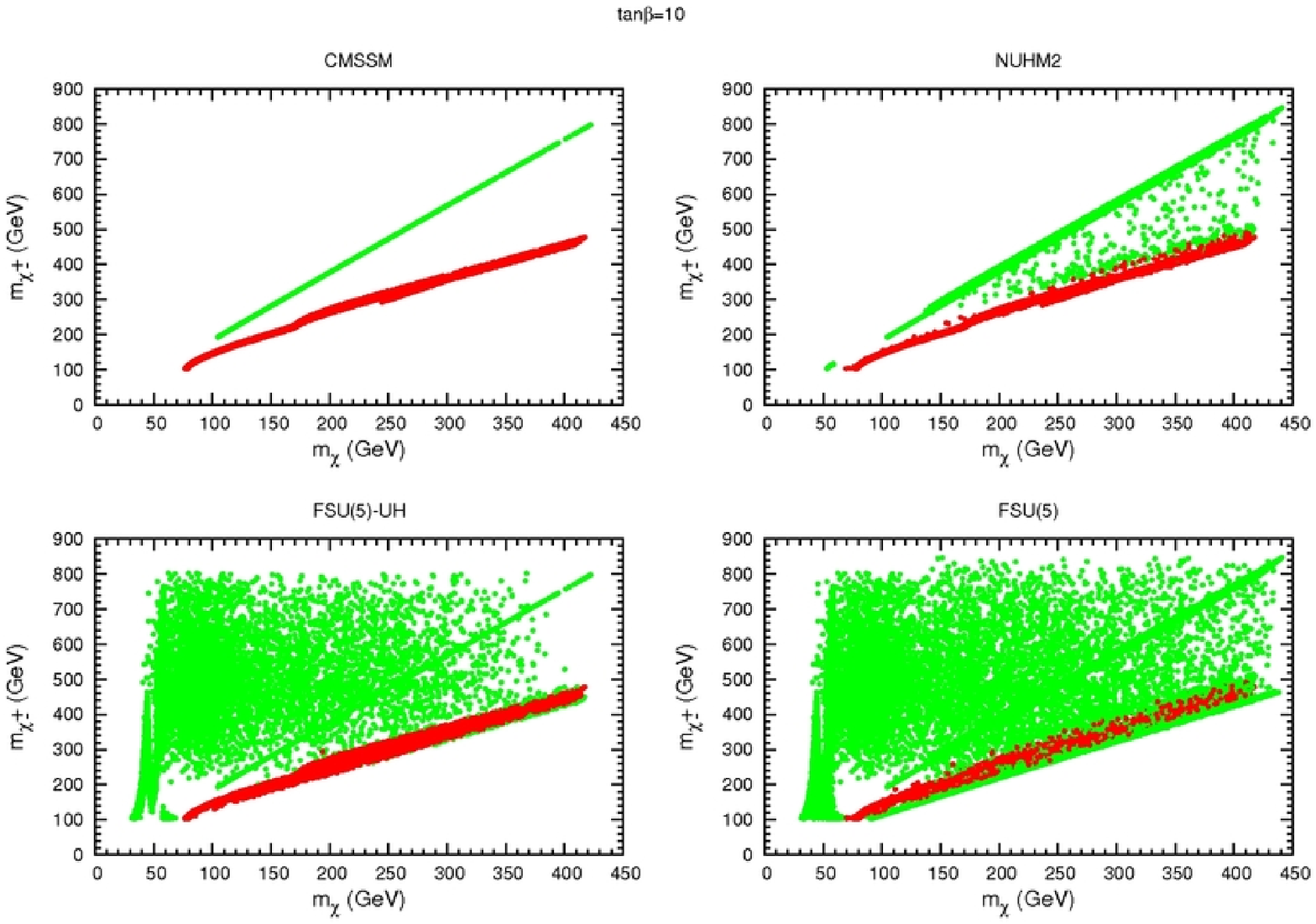}
\caption{Plot in the $m_{{\chi^{\pm}_1}} - m_{\chi}$ plane. Color coding same as in Fig. \ref{fig:7}
\label{fig:11}}
\end{figure*}

In Figures  \ref{fig:8}, \ref{fig:9}, \ref{fig:10} and \ref{fig:11} we show the results in $m_A$ vs. $m_{\chi}$,  $m_{\tilde{\tau}}$ vs. $m_{\chi}$, $m_{\tilde{t}}$ vs. $m_{\chi}$ and $m_{{\chi^{\pm}_1}}$ vs. $m_{\chi}$ planes respectively. The color coding in these figures is the same as for Fig.   \ref{fig:7}.  We can see  that the FSU(5) model gives rise to signals, corresponding to a relatively light
$\tilde{\tau}$ and $m_A$, which can be seen by the IceCube/DeepCore experiment. This is not the case for CMSSM and NUHM2 models.

For ${\rm tan\beta}=10$ the lower mass bounds on the lightest neutralino for the four models (CMSSM, NUHM2, FSU(5)-UH and FSU(5)) are 76.7 GeV, 53.1 GeV, 32.2 GeV and 31.6 GeV respectively. This is consistent with a recent study \cite{Vasquez:2010ru} which found a  lower bound of 28 GeV on the mass of the LSP neutralino.
 From Fig \ref{fig:4} we see that this neutralino in the CMSSM case comes from the focus point region and when the Bino or Higgsino mixing is large. For NUHM2, it is from the h-resonance channel, whereas for FSU(5)-UH it is from the "non-identified" region. We note that the muon flux is highest when the LSP dark matter neutralino is mainly Bino-Higgsino like, and this observation is valid for the CMSSM as well as for its extensions.

  Note that the IceCube/DeepCore bound is a conservative one and the muon flux limits can be improved by an order magnitude  \cite{Ellis:2009ka}. As noted in \cite{Ellis:2009ka} their are prospects of detecting the CMSSM focus point (FP) region in the IceCube/DeepCore experiment. This can also be seen in Figures \ref{fig:3}-\ref{fig:5} for the CMSSM. Going from CMSSM to FSU(5) changes this, and in addition to more points in the FP region we also have some points from the non-identified and h-resonance regions. The non-universality of the Higgs mass parameters opens up the A-funnel region where resonant annihilation occurs through the CP odd Higgs boson A. The Bino-Wino coannihilation \cite{Baer:2005jq} channel arises from the gaugino non-universality in Eq. (\ref{gauginoCondition}) and occurs for $2 M_2 \sim M_1$ at $M_{\rm GUT}$. As we can see from the muon flux plots, this region of
 the parameter space is not detectable with the current IceCube/DeepCore experiment.

     Finally, in TABLE \ref{table1} we present three FSU(5) benchmark points for ${\rm tan\beta}=10,30,50$ which yield observable muon fluxes. The first point belongs to the stau coannihilation region, the second point is associated with the Bino-Higgsino dark matter with light charginos, and the third corresponds to the h-resonance region.


\section{Conclusions\label{conclusions}}
We have considered indirect neutralino dark matter detection in two sets of supersymmetric flipped SU(5) models. These two sets of models have non-universal soft gaugino masses at $M_{GUT}$ that are related by the underlying $SU(5)\times U(1)_X$ gauge symmetry. The supersymmetry breaking soft Higgs $\rm {masses}^2$, associated with $H_u$ and $H_d$, are equal at $M_{GUT}$ in one set of models (FSU(5)-UH) but not in the other (FSU(5)). We have provided estimates of the flux, from annihilating neutralinos in the Sun, of neutrino induced muons, and considered prospects of detecting this flux in the IceCube/DeepCore detector. Some uncertainties arise in converting the muon flux into spin dependent neutralino-nucleon cross sections that we have briefly discussed. We offer comparisons with previously studied CMSSM and NUHM2 models, and also highlight some benchmark models in flipped SU(5) with varying neutralino compositions which can be tested by the IceCube/DeepCore experiment.
Our results for NUHM2, FSU(5)-UH and FSU(5) show more points above the projected IceCube limit compared to the CMSSM, and hence a greater prospect of detection. This is to be expected because the models are less constrained than the CMSSM and possess additional free parameters.

%
\acknowledgments
We thank Hasan Y$\ddot{\rm u}$ksel for his helpful advice, discussions and collaboration in
the early stages of this work. We are also grateful to Joakim Edsj$\ddot{\rm o}$, Gustav Wikstr$\ddot{\rm o}$m and Stephano Profumo for discussions related to Darksusy and our results.  We also thank Rizwan Khalid and Shabbar Raza Rizvi for very useful comments and suggestions. This work
is supported in part by the DOE Grant No. DE-FG02-91ER40626
(M.A., I.G.,  and Q.S.) and GNSF Grant No. 07\_462\_4-270 (I.G.).

%


\newpage

%
%
%
%
%

%
%

\begin{table}[t!]
\centering
\begin{tabular}{lccccc}
\hline
\hline
                & Point 1 & Point 2    & Point 3                  \\
\hline
$m_{0}$          &  107       & 1349    & 1335                      \\
$M_{1} $         &  691       & 295     & 176                       \\
$M_{2} $         &  607       & 848     & 519                       \\
$M' $            &  695       & 272     & 161                       \\
${\rm tan\beta}$     &  10        & 30      & 50                        \\
$A_0$        &  0         & 0       & 0                         \\
$\mu$            &  372       & 115     & 120                       \\
$m_{A}$          &  965       & 616     & 731                       \\

\hline
$m_h$            &  115       & 118     & 115                     \\
$m_H$            &  971       & 620     & 736                     \\
$m_A$            &  965       & 616     & 731                    \\
$m_{H^{\pm}}$    &  974       & 626     & 742                    \\

\hline
$m_{\tilde{\chi}^0_{1,2}}$
                 &  275,361    & 86,125  & 58,128               \\
$m_{\tilde{\chi}^0_{3,4}}$
                 &  380,514 & 148,696 & 132,439           \\

$m_{\tilde{\chi}^{\pm}_{1,2}}$
                 & 356,509  & 120,686  & 432,120          \\
$m_{\tilde{g}}$  & 1369    &   1943     & 1263                \\

\hline $m_{ \tilde{u}_{L,R}}$
                 & 1264,1220  & 2178,2146  & 1709,1683         \\
$m_{\tilde{t}_{1,2}}$
                 & 900,1172   & 1493,1872  & 1124,1325           \\
\hline $m_{ \tilde{d}_{L,R}}$
                 & 1267,1214 & 2179,2118   & 1711,1687          \\
$m_{\tilde{b}_{1,2}}$
                 & 1140,1205 & 1852,2014   & 1301,1368            \\
\hline
$m_{\tilde{\nu}_{1}}$
                 & 417       &  1473     &  1370                     \\
$m_{\tilde{\nu}_{3}}$
                 & 410       &  1433     &  1195                    \\
\hline
$m_{ \tilde{e}_{L,R}}$
                &  428,284   &  1476,1305 & 1373,1338            \\
$m_{\tilde{\tau}_{1,2}}$
                &  277,422   &  1211,1435  & 941,1196             \\
\hline

$\sigma_{SI}({\rm pb})$
                & $1.8\times 10^{-8}$ & $7.65\times 10^{-8}$ & $5.7\times 10^{-8}$
                              \\

$\sigma_{SD}({\rm pb})$
                & $3.9 \times 10^{-5}$ & $9.75 \times 10^{-4}$ & $8.0\times 10^{-4}$
                    \\

$\Omega_{CDM}h^2$
                & 0.075       & 0.077   &  0.093            \\

$\phi_{\mu+\overline{\mu}}(\rm{km^{-2}yr^{-1}})$
                & 118       & 3728   &  1170            \\

\hline
\hline
\end{tabular}
\caption{ Sparticle and Higgs masses (in GeV),
with $m_t=173.1\, {\rm GeV}$. These benchmark points satisfy all the constraints
imposed in Section~ \ref{sec:results}.
\label{table1}}
\end{table}

\end{document}